\documentclass[11pt,twoside,usenatbib]{article}

\usepackage{asp2006}
\usepackage{epsfig}
\usepackage{graphicx}
\usepackage{lscape}
\usepackage{natbib}
\bibstyle{apj}
 \newcommand{\Mdot}{\dot{M}}
\newcommand{\Msun}{M_{\odot}} 

\begin{document}
\title{Embedded, Accreting Disks in Massive Star Formation}   %%% Fill in title
\author{Kaitlin M. Kratter and Christopher D. Matzner}
\affil{Department of Astronomy and Astrophysics, University of Toronto, Toronto, ON M5S
3H4}

\author{Mark R. Krumholz\altaffilmark{1}} \affil{Department of
Astrophysical Sciences, Princeton University, Princeton, NJ
08544-1001}
\altaffiltext{1}{Hubble Fellow}
\begin{abstract} %%% Abstract to run on from here.
Recent advances in our understanding of massive star formation have made clear the important role of protostellar disks in mediating accretion. Here we describe a simple, semi-analytic model for young, deeply embedded, massive accretion disks. Our approach enables us to sample a wide parameter space of stellar mass and environmental variables, providing a means to make predictions for a variety of sources that next generation telescopes like ALMA and the EVLA will observe. Moreover we include, at least approximately, multiple mechanisms for angular momentum transport, a comprehensive model for disk heating and cooling, and a realistic estimate for the angular momentum in the gas reservoir. We make predictions for the typical sizes, masses, and temperatures of the disks, and describe the role of gravitational instabilities in determining the binarity fraction and upper mass cut-off. We also address the relationship between cooling time and fragmentation seen in a range of disk simulations. 
\end{abstract}

\section{Introduction}
Massive star formation is finally coming into focus in our telescopes and on our computer screens. Increasing telescope sensitivity and resolution have provided our first glimpse of deeply embedded, young sources, and ever-expanding computational power has enabled us to numerically probe the scales on which disks form. However, we are still unable to survey the parameter space of massive star formation numerically due to finite computational resources, and observationally due to a lack of statistics. Analytic work can fill this crucial hole. In the coming years, our ability to search for these disks will improve dramatically, and so we aim to survey the characteristics of massive disks as a function of stellar mass and environment in order to aid in their identification and test current theoretical paradigms.

The presence of disks has been confirmed observationally for stars as massive as early B type \citep{cesaroni07a,2007arXiv0707.1279Z}. Disks are most likely present around O stars as well, as the lack of evidence is attributable to selection effect on two fronts: (1) the timescales on which these disks exist are short, perhaps only marginally longer than the accretion time (Yorke, 2007 these proceedings), and (2) the disks are deeply embedded in high column cores for the majority of their lifetime, making them difficult to observe.

Disks are a necessary consequence of angular momentum conservation in any model for high mass star formation, and play a crucial role in overcoming the radiation pressure barrier \citep{WC1987,YS2002,2005ApJ...618L..33K}. But despite the increase in disk detections, we are still ignorant as to how disks differ across the parameter space of stellar mass and time. To begin to answer these questions, we have constructed one-zone semi-analytic models. 

Our approach is unique in that we define an accretion model for the disk, which is a fit to a range of simulations, and couple this to a heating and cooling model. As a result, we do not impose an artificial steady-state to derive accretion rates. This enables us to self-consistently track the disk's dynamical evolution through a series of thermal equilibria.

In order to derive results across this parameter space,  we take the simplified case of a single zone disk. While this limits the details of our model in some sense, it allows us to be more thorough in other respects. We account for the dependence of gravitational torques on both disk-to-total mass ratio and on Toomre's  Q parameter. In addition, we allow the disks to fragment both locally and globally when sufficiently unstable. Moreover, we consider fluctuations of the vector angular momentum in the infall due to realistic turbulence in the collapsing core. Finally, we employ a sophisticated model for the irradiation of the disk midplane, which is crucial to correctly predicting the thermal and dynamical evolution of the disk.  With this model in hand, we can evolve star-disk systems across a wide mass range in various star-forming environments. For a more detailed description of the model and results, see \cite{KMK2007}.

\section{Model properties}
To construct a self-consistent model for evolving disks we need three major components: a model for how the system receives material from its environment, a model for how the disk transports mass and angular momentum, and a model for the thermal physics. 
We divide the elements of our model into two categories: imposed external parameters, and induced internal ones, set by the physics operating within the disk. We begin by discussing the former. The most relevant imposed parameters are our model for the accretion rate, $\Mdot(t)$, and the angular momentum infall rate, $\dot{J}(t)$, for these set the fate of the star-disk system. We choose to employ a \cite{MT2003} core model, though an arbitrary formation model can be subsituted. This model, for a given core mass, temperature, and star formation efficiency, $\varepsilon$, gives the accretion rate onto the star disk system as a function of time. Two other global parameters influence the accretion rate of matter and angular momentum in this model: initial core temperature, and core surface pressure. These two parameters convey the type of star forming environment. While field stars are likely borne of cooler, less dense clouds, clustered stars should arise from a hotter, high pressure medium.

\subsection{Angular momentum infall}
Star forming cores are often modeled as solid body rotators characterized by the ratio $\beta$ of rotational to gravitational energy. While observations \citep{1993ApJ...406..528G} can constrain velocities, theoretically there is no compelling reason to use these to infer solid body rotation. However, turbulence in cores can also serve as an angular momentum reservoir. In fact, observations are equally consistent with a model in which core motions are purely turbulent. \citep{2000ApJ...543..822B}. We therefore use the \cite{MT2003} models to determine the magnitude of the turbulent velocities needed for the cores to be in rough hydrostatic balance with the surrounding medium--massive cores are not thermally supported--and calculate the net angular momentum on radial shells from a simulated velocity field.

\subsection{Accretion Model and Thermal Physics}
Given a model for how material and angular momentum fall onto the disk, we must determine how these are processed. We include two mechanisms for angular momentum and mass transport in the disk: gravitational instability driven turbulence (GI) and magnetorotational instability driven turbulence  (MRI) . By fitting data from numerical simulations  across our parameter space  we derive a composite accretion rate  based on two parameters: Toomre's Q =  $\frac{c_s \kappa} {\pi G \Sigma}$ \citep{Toom1964}, and $\mu =$$\frac{M_d }{M_d +M_*}$. Using a fit to a range of simulations, we compute an effective \cite{SS1973}  $\alpha$ viscosity as a function of these disk variables at every timestep. We equate $\alpha$ with a mass accretion rate using the steady-state formula: $\Mdot = $$3 \alpha c_s^3 \over {GQ}$. 

 Based on the relative strength of gravitational versus magnetic instabilities, we take the simplifying assumption of a constant MRI viscosity $\alpha_{MRI} = 0.01$, a reasonable estimate for its strength based on simulations such as \cite{2004ApJ...616..357F}. For the gravitational instability we use data from three sources. First, we use \cite{1996ApJ...456..279L}, who conduct 2D numerical simulations of massive protostellar disks. They span a range of masses as well as disk temperatures, however they suppress local instability by using an adiabatic equation of state. These are complemented by the results of \cite{Gam2001}, who examines the purely local case in the shearing sheet approximation. Finally we use the global SPH simulations of \cite{2003MNRAS.339.1025R}, \cite{Lod04}, and \cite{LodRi05},  who find that $\alpha$ limits to a maximum value as $Q$ approaches unity. This value depends on the rate at which the disk cools.
 \begin{figure}
\begin{center}
%\epsscale{0.7}
%\plotone{accmod.eps}
%\epsscale{0.75}
\includegraphics[scale=0.5]{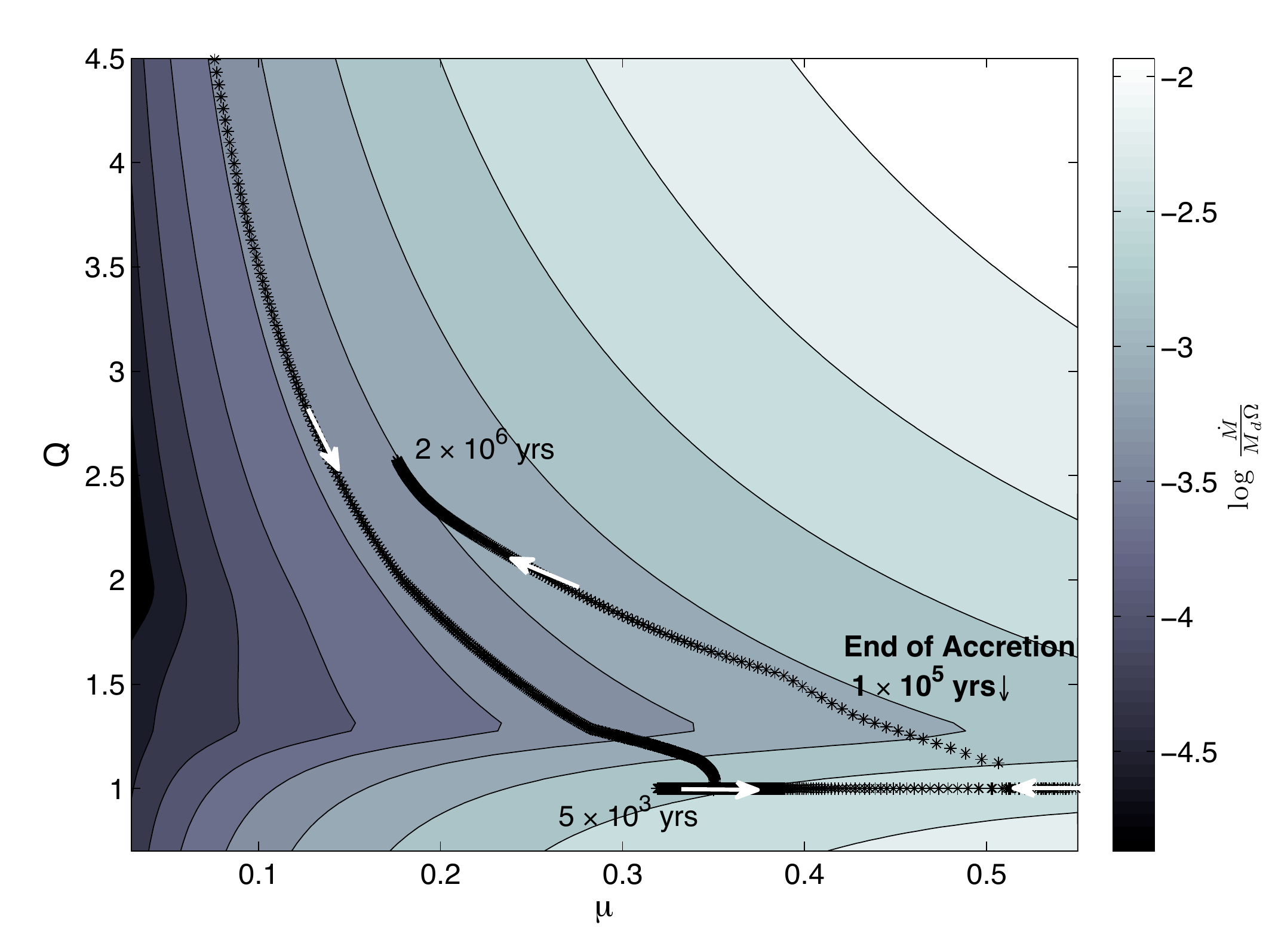}
\caption{{Evolutionary track in the $Q, \mu$ plane of a 15$\Msun$ star-disk system overlayed on the contours of our accretion model (contour spacing is 0.2 dex). The white arrows superposed on the track show the direction of evolution in time with several key times marked.}}
\label{AccModel}
\end{center}
\end{figure}

The background of Figure \ref{AccModel} shows contours of constant $\Mdot$ in our fiducial accretion model. For a complete discussion of the model and its correspondence with these simulations, see \cite{KMK2007}.  The model shown here includes a background $\alpha_{MRI}$ active at all $Q$ and $\mu$ values, and two components to the GI driven transport: global spiral arm torques, which we fit with a model that goes to zero as $Q \rightarrow2$ \citep{2006MNRAS.365.1007G}, and a local gravito-turbulence component that begins to dominate when $Q\leq 1.3$:
\begin{eqnarray}
\alpha_{\rm loc} &=& \max\left[0.14(\frac{1.3^2}{Q^2} -1)(1-\mu)^{1.15},0\right] \\
\alpha_{\rm glob}&=& \max\left[1.4\times 10^{-3}\mu^{-5/4}\frac{2-Q}{Q^{1/2}},0\right].
%\alpha_tot = \sqrt({\alpha_{\rm loc}^2 + \alpha_{\rm glob}^2 + \alpha_{MRI}^2})
\end{eqnarray}
 The former scaling is equivalent to the analytic formula derived by  \cite{1990ApJ...358..515L}  (eq. 16) modified by a mild $\mu$ dependence, comparable to the scale height dependence derived in equation (2.5) of \cite{1987MNRAS.225..607L}. The effective total $\alpha$ we use is simply $\alpha_{\rm tot} = \sqrt{\alpha_{\rm MRI}^2+\alpha_{\rm {loc}}^2+\alpha_{\rm glob}^2}$. Our accretion model is only as accurate as the data on which it is based, and as more simulations become available that map the effects of gravitational instability in different regimes, we expect to improve this fit.  We justify the division of the gravitational instability into two components below. 
 
 We bound our model in the $Q - \mu$ plane by two types of disk fragmentation. If $\mu $ exceeds 0.5, i.e. the disk mass exceeds the star mass, we put the excess mass and angular momentum into a binary. If $Q$ drops below unity, mass from the disk is turned into bound fragments that accrete with the disk, but do not contribute to the disk surface density. The first type of fragmentation is based on simple stability arguments: such a disk is no longer Keplerian, and is unlikely to remain stable over many orbital timescales. We discuss the $Q$-based fragmentation prescription below.

\subsubsection{Global vs Local Instability}
In order to better parameterize the way that disks accrete as a function of their physical state, we consider the gravitational instability in two different regimes, ''global'' and ''local". Here global and local refer to the scale of the instability itself. These can roughly be thought of as describing whether or not the wavelength of the instability is comparable to the disk radius, or the disk scale height. Alternatively one can express this in terms of the azimuthal mode number, $m$ which characterizes the unstable mode. \cite{Lod04} and (2005) have convincingly shown that one can distinguish between these two regimes, and that there appears to be a dependence on the mass of the disk relative to the star \citep{2003MNRAS.339.1025R}. As disks approach the $Q=1$ boundary, more massive disks show more strength in lower $m$-modes as compared to light disks, which show power evenly distributed in a range of modes from $m=2-12$. This division can be at least partly attributed to the difference in aspect ratio between low and high $\mu$ disks at the same $Q$. Our accretion model demonstrates that these two regimes have different dependences on the state of the disk: the excitation of low $m$-modes seems to depend preferentially on the parameter $\mu$ above-- while the high $m$ modes depend primarily on the classic Toomre instability.

\subsubsection{Fragmentation} \label{Fragmentation}
Cooling times also play a role in determining whether or not a violently unstable disk remains self-regulated by turbulence or begins to fragment \citep{2003MNRAS.339.1025R}.  These authors find that a slowly cooling disk can sit \emph{at} $Q=1$ in a steady state, however for shorter cooling times, $Q$ drops below unity, and fragmentation occurs. Based on the simulations of \cite{2003MNRAS.339.1025R} and \cite{Lod04} and (2005), we argue that cooling time regulates the value of $Q$ at which the disk reaches an equilibrium state: if this values is below unity, the disk fragments. For example, \cite{2003MNRAS.339.1025R} note that a sufficiently slowly cooling disk reaches an equilibrium value at a value \emph{higher} than $Q=1$, suggesting that the cooling time sets the level of turbulence, and thus temperature where a disk equilibrates.  Because we have combined a realistic heating model with this accretion model, we can self-consistently determine whether or not the disk fragments. If a disk drops below $Q=1$, its effective cooling time is too short to remain self-regulated.

\begin{figure}
\begin{center}
\plottwo{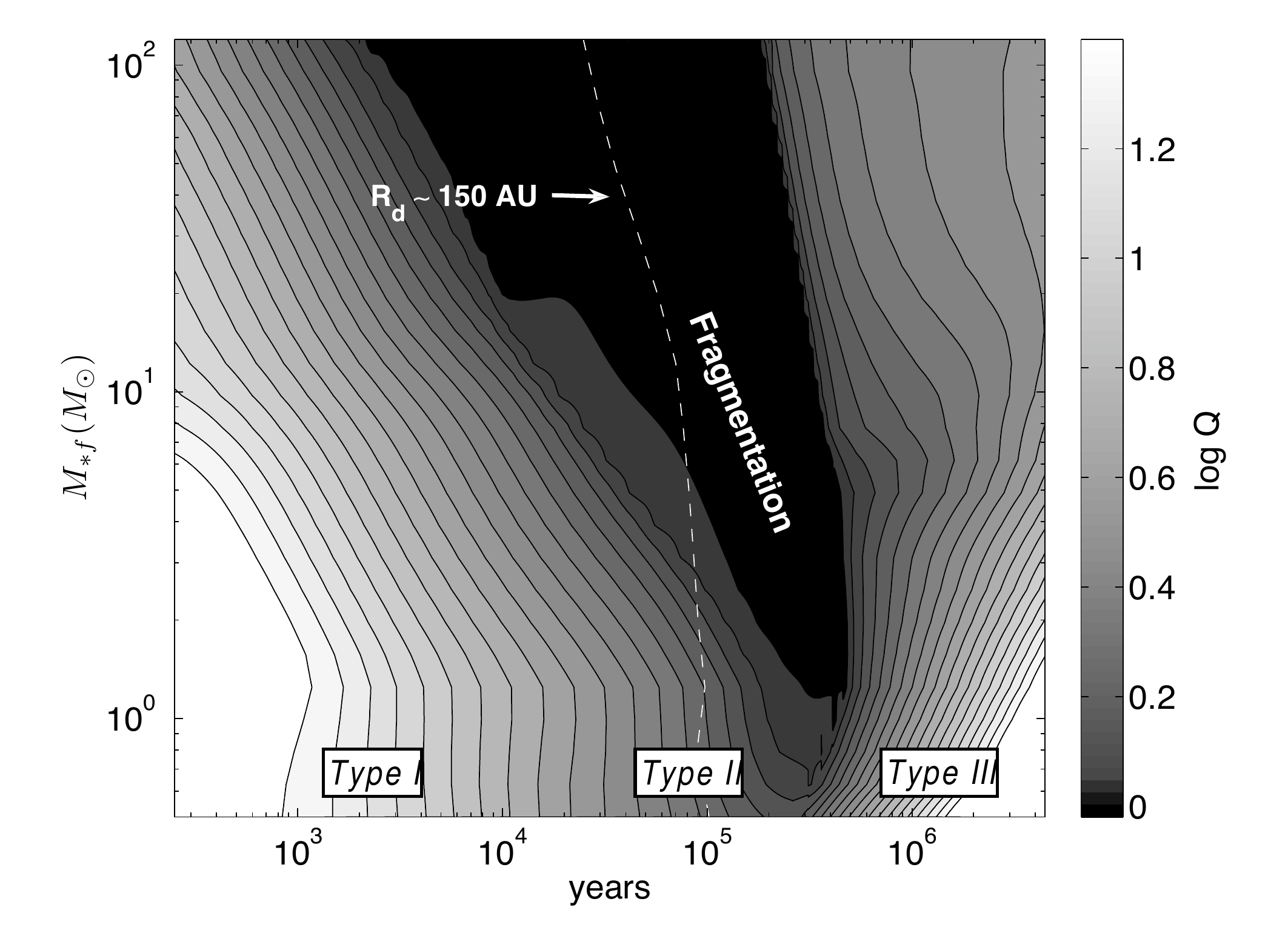}{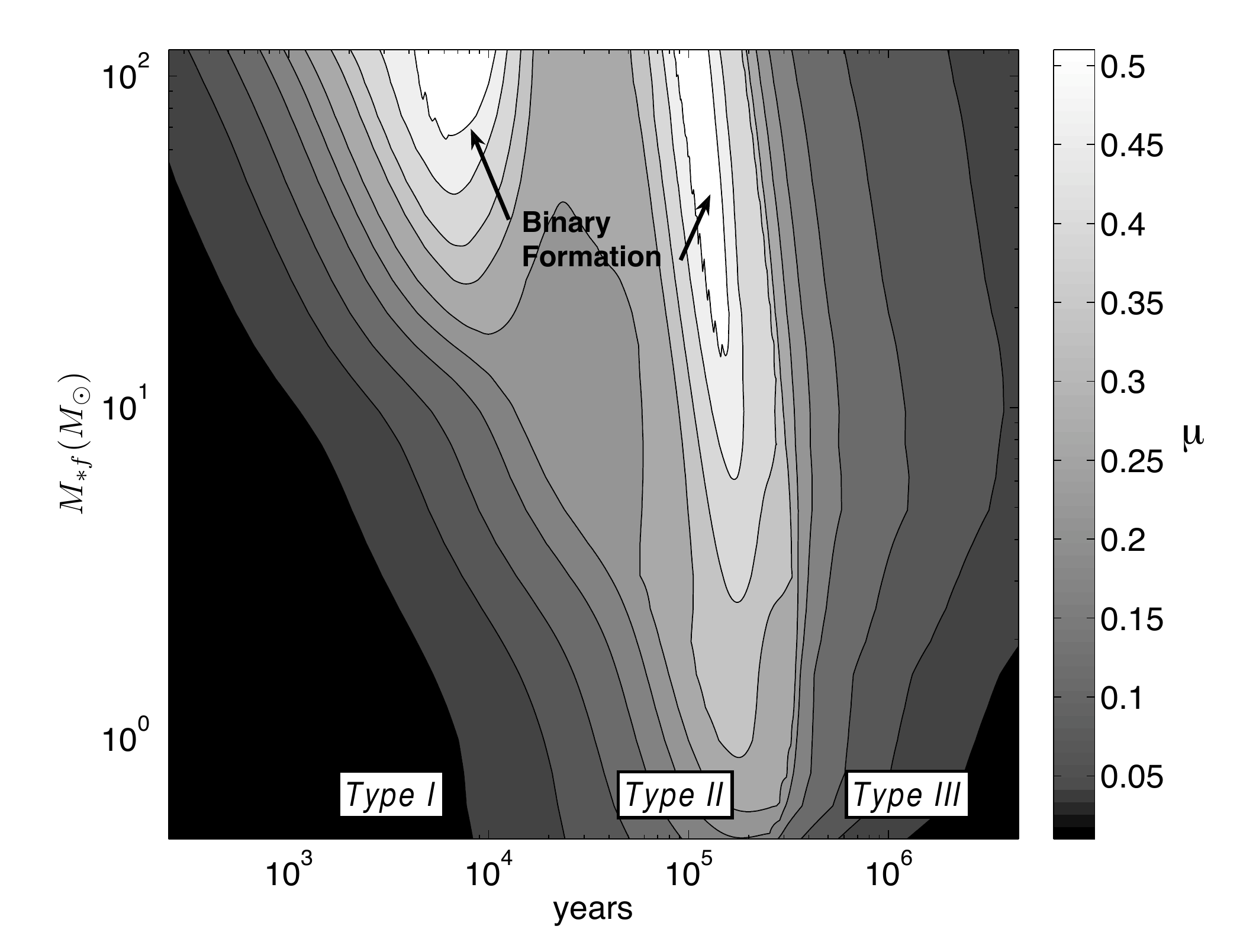} 
\caption{{Contours of log$(Q)$ (left) and $\mu$ (right) over the accretion history of a range of stellar masses for the fiducial model sequence. Contours are spaced by .1 dex (left) and .05 (right). Contours of Q show that at low final stellar masses, disks remain stable against the local instability throughout accretion, while higher masses, all undergo a phase of local instability. We overplot the typical radius at which instability sets in \citep{KM06}. The second plot illustrates the frequency of binary formation in disks around more massive stars. See \S \ref{results} for a discussion of the three labeled disk types.}}
\label{contours}
\end{center}
\end{figure}

 \section{Results} \label{results}
 Our results are best summarized by Figures 1 and 2, and from these we make rough predictions for the characteristics of massive disks. In Figure 1 we show the evolutionary track of a massive star accreting from a $30\Msun$ core, with an accretion efficiency of $50\%$. The track is superposed on a contour plot of our model for mass transport within the disk as a function of $Q$ and $\mu$. There are several noteworthy features: the first is that the system spends most of its time in the regime where gravitational rather than magnetic instabilities dominate accretion, implying that our cursory treatment of magnetic effects, at least in the dense disk stages, is reasonable. Second, the disk accretes from the core more rapidly than it can deposit material onto the star, causing the disk mass to build up. Once the disk drops below $Q=1$, we assume that the local instability has saturated, matter starts to condense out of the disk and turn into fragments that no longer contribute to the surface density of the disk. Because of the rapid accretion rates, however, the disk mass continues to build up, eventually reaching the binary formation barrier of $\mu = 0.5$. As accretion shuts off from the core, the disk drains on to the star, and $Q$ begins to rise. 
 
 Figure 2 maps out the evolution of $Q$ and $\mu$ across a broad mass range. One can see three roughly distinct periods marked as types I-III: Type I systems have low disk masses and are stable. Type II disks are massive, on average $35\%$ of the total system mass,  show strong spiral structure, and may be fragmenting. Disk radii during this stage are of order a couple hundred AU, with outer disk temperatures of roughly $100K$.  As accretion from the core shuts off, disks enter the Type III stage where they are once again stable and low mass, but have significantly larger radii than during the Type I stage due to the high-$j$ material deposited onto the disk at the end of accretion.
  
 \subsection{Environmental Effects}
 We explore the effect of varying the star-forming environment by varying the core binding pressure, core temperature, core angular momentum, and strength of the MRI over an order of magnitude. For massive stars, neither binding pressure nor core temperature have dramatic effects on the evolution of systems. The strength of the MRI turbulence influences the maximum disk mass a system attains by changing the amount of matter the disk processes at early times when it is gravitationally stable. This has implications for binary formation: weaker MRI increases the likelihood of forming a binary by forcing the disk mass up at early times. Core angular momentum has the strongest effect on the systems' evolution.  Using several realizations of a turbulent velocity field with the same normalization and power spectrum, the disk radius can vary by a factor of a few. Such fluctuations carry through to other calculated quantities like disk temperatures, densities, and binary frequency. This is especially true in a one-zone model where all quanties are calculated at the outer radius, which is set by the angular momentum in the core; in reality the effect on the evolution of the disk as a whole may be slightly less significant.
 
\subsection{Binary Formation and Upper Mass Limits}
Figure 2(b) illustrates the ubiquity of disk-born binaries. Although we do not have a good way to determine the resultant mass ratio, or whether a larger multiple system is forming, due to large-scale disk instability it seems likely that cores above roughly $20\Msun$ will form a multiple through disk fragmentation. Because higher mass systems tend to form binaries earlier in their evolution, this provides a natural pathway to equal mass binaries. As pointed out by \cite{krumholz07c}, as long as the two pre-main sequence stars are reasonably close, which should happen through disk formation when the disk is quite small, they will undergo mass transfer and grow simultaneously. Binaries that form later, and likely at larger separations, may have smaller mass reservoirs and thus lower masses. Surveys reviewed by Gies (2007, these proceedings) have suggested a trend towards higher mass ratios at larger separations. Although fragmentation due to the local instability may also be a mechanism for binary formation, the fate of such fragments is unclear \citep{2007ApJ...665..478K,Lod04}  

While we cannot follow the fragments' evolution in analytic models, there are other consequences of fragmentation. \cite{KM06} point out that disks become catastrophically unstable above a critical accretion rate. Looking at the upper region of Figure 2 (a) one can see that the period of fragmentation lasts for most of the accretion history of the system. If in fact the disk is consumed by fragmentation, the central star system may be starved of disk material. This can set an upper limit of around $150\Msun$ on the mass to which a star can accrete through a disk \citep{KM06}. The formation of binaries may also play a role in limiting the upper mass of a single star, as wholescale fragmentation becomes increasingly likely with increasing specific angular momentum in the gas reservoir. The exact scaling between mass and angular momentum may be more complicated if the entire reservoir is not initially bound to the core.

\section{Summary}
We have constructed a simple, semi-analytic evolutionary model for rapidly accreting, massive, protostellar disks. We use this to predict the typical characteristics of disks as a function of stellar mass and environment. Our approach is novel in that we create a dynamical accretion model and a thermal model, which we couple to solve for the state of the disk self-consistently as a function of time. By accounting for the feedback loop between the rate at which the disk accretes, and the disk temperature, we do not need to rely on steady-state assumptions or assume an ideal equation of state for the disk gas. Rather we create a detailed heating and cooling model including viscous dissipation from accretion, radiation from central star (calculated based on its mass, evolutionary phase, and accretion rate), and cooling in the optically thin and thick regimes. We find that disk are not in a steady-state with a constant accretion rate and disk-to-star mass ratio, but evolve through three distinct phases. They start out at low masses, and high $Q$ values, but they evolve into a very unstable phase where disks become gravitationally unstable, show strong spiral structure and fragment both locally and globally. Finally, once the mass reservoir has been exhausted, disks continue to drain onto the central star, declining in mass but increasing in radius due to viscous spreading.

These models can provide baseline predictions for future observations, and can be updated as new simulation data becomes available. Moreover, because of the separation between the disk accretion rate and mass infall rate, this model can be applied to a wide range of star formation scenarios across various mass ranges and environments.

  %\section{}   %%% Top level section head (remove "%" symbol)
%\subsection{}   %%% Second level section head (remove "%" symbol)
%\subsubsection{}   %%% Lowest level section head (remove "%" symbol)
%\section*{}    %%% Unnumbered top level section head (remove "%" symbol)
%\subsection*{}   %%% Unnumbered second level section head (remove "%" symbol)

\acknowledgements %%% Text of acknowledgements runs on after this command.
We are grateful to the SOC and LOC for their hard work in organizing and running an excellent, productive meeting. We would also like to thank G. Lodato and W. K. Rice for helpful discussions regarding their simulations. MRK is supported through Hubble Fellowship grant \#HSF-HF-01186 awarded by the Space Telescope Science Institute, which is operated by the Association of Universities for Research in Astronomy, Inc., for NASA, under contract NAS 5-26555. This research was supported in part by the National Science Foundation under Grant No. PHY05-51164.
%\bibliography{diskbibU}
%%% THE BIBLIOGRAPHY
%%%
%%% CONSULT SECTION 3 OF "INSTRUCTIONS FOR AUTHORS" FOR HOW TO USE NATBIB.
%%% AUTHORS ARE ENCOURAGED TO USE EITHER THE "THEBIBLIOGRAPY" ENVIRONMENT
%%% BY UNCOMMENTING (DELETING THE "%" SYMBOL) THE COMMANDS BELOW, OR BY
%%% USING THE BIBTEX ENVIRONMENT. TO FIND OUT WHICH IS APPLICABLE TO YOUR
%%% CONTRIBUTION, CONSULT THE VOLUME EDITORS FOR YOUR PROCEEDINGS.
%%%

%\begin{thebibliography}{}
%\bibitem[]{}
%\bibitem[]{}
%\bibitem[]{}
%\bibitem[]{}
%\bibitem[]{}
%\bibitem[]{}
%\bibitem[]{}
%\bibitem[]{}
%\bibitem[]{}
%\bibitem[]{}
%\bibitem[]{}
%\bibitem[]{}
%\end{thebibliography}

\end{document}